\documentstyle[aps]{revtex} \begin{document}

\draft
\title{ Decoherence in Bose-Einstein Condensates:  \\            
        Towards Bigger and Better Schr\"odinger Cats }
\author{ Diego A. R. Dalvit$^1$
	   \thanks{e-mail: \tt dalvit@lanl.gov}, 
         Jacek Dziarmaga$^{1,2}$
         \thanks{ e-mail: \tt dziarmaga@t6-serv.lanl.gov }, 
         and 
         Wojciech H. Zurek$^1$ \thanks{e-mail: \tt whz@lanl.gov}}
\address{ 1) Los Alamos National Laboratory, T-6, Theoretical Division, 
             MS-B288, Los Alamos, NM 87545, USA                              \\
          2) Institute of Physics, Jagiellonian University, Krak\'ow, Poland }

\date{\today}
\maketitle
\tighten

\begin{abstract} 
{\bf We consider a quantum superposition of Bose-Einstein condensates in
two immiscible internal states. A decoherence rate for the resulting 
Schr\"odinger cat is calculated and shown to be a
significant threat to this macroscopic quantum superposition of BEC's. An
experimental scenario is outlined where the decoherence rate due to the
thermal cloud is dramatically reduced thanks to trap engineering and
``symmetrization" of the environment which allow the Schr\"odinger cat to be an
approximate pointer state. }

\end{abstract}

\section{ Motivation and Summary of Results}

Microscopic quantum superpositions are an everyday physicist's
experience.  Macroscopic quantum superpositions, despite nearly a century
of experimentation with quantum mechanics, are still encountered only very
rarely. Fast decoherence of macroscopically distinct states is to be
blamed \cite{zurek}. In spite of that, recent years were a witness to an
interesting quantum optics experiment \cite{paris} on decoherence of 
a few photon
superpositions. Moreover, matter-wave interference in fullerene
$\rm{C}_{60}$ has been observed \cite{insbruck}. Another recent
experiment has succeeded in ``engineering'' the environment in the context
of trapped ions \cite{wineland}. This success plus rapid progress in
Bose-Einstein condensation (BEC) of alkali metal atomic vapors \cite{bec}
tempt one to push similar investigations even further into the macroscopic
domain. The condensates can contain up to $10^7$ atoms in the same quantum
state. What is more, it is possible to prepare condensates in two
different internal states of the atoms. Some of these pairs of internal
levels are immiscible, and their condensates tend to phase separate into
distinct domains with definite internal states \cite{spinor}. The
immiscibility seems to be a prerequisite to prepare a quantum
superposition in which all atoms are in one or 
the other internal state, 
$|\psi\rangle = (|N,0\rangle + |0,N\rangle )/\sqrt{2}$, 
where $N$ is the total number of condensed atoms. There are at least
two theoretical proposals how to prepare a macroscopic quantum
superposition in this framework \cite{lz,gs}.  Neither of them addresses
the crucial question of decoherence.

We find the quantum superposition state involving significant number of
BEC atoms to be practically impossible in
the standard harmonic trap. Our master equation (derived in Section IV),
when applied to the standard harmonic trap of frequency $\omega$, gives
for the decoherence rate due to the environment of noncondensed atoms

\begin{equation}\label{dr}
t^{-1}_{\rm dec} 
\;\approx\; 
16 \pi^3\; \left( 4\pi a^2 \frac{N_{\rm E}}{V} v_T  \right) \; N^2   \;\;,
\end{equation}
where $N$ is a number of condensed atoms, $v_T=\sqrt{2k_{\rm B}T/m}$ is a
thermal velocity in the noncondensed thermal cloud, $a$ is a scattering
length, $V$ is a volume of the trap, and $N_{\rm E}$ 
is a number of atoms in the thermal cloud,

\begin{equation}\label{NE}
N_{\rm E}
\;\approx\;
e^{\frac{\mu}{k_{\rm B} T}} \;
\left( \frac{k_{\rm B}T}{\hbar\omega} \right)^3 \;\;.               
\end{equation}
Here $\mu<0$ is a chemical potential. 

Equation (\ref{dr}) gives the rate of decoherence to the leading
order in the fugacity $z=\exp(\beta\mu)$ and also to the leading order in
the condensate size, $N$. Next-to-leading order terms are $O(z^2)$ and
$O(N)$, in agreement with Refs.\cite{anglin,wallsroul}, so they were 
neglected here. We also assume
that $N$ is small enough for a condensate to live in a single mode as
opposed to the large $N$ Thomas-Fermi limit. Given all these assumptions,
in the standard harmonic trap setting Eq.(\ref{dr}) is a lower estimate
for the decoherence rate.

Even without going into details of our derivation, which are given in
Section IV, it is easy to understand where a formula like Eq.(\ref{dr})
comes from. $N^2$ is the main factor which makes the decoherence rate
large. It comes from the master equation of the Bloch-Lindblad form
$\dot\rho\sim -[N_{\rm A}-N_{\rm B},[N_{\rm A}-N_{\rm B},\rho]]$, with
$\rm A$ and $\rm B$ the two internal states of the atoms. $N^2$ is the
distance squared between macroscopically different components of the
superposition $(|N,0\rangle + |0,N\rangle)/\sqrt{2}$ - 
the common wisdom reason why
macroscopic objects are classical \cite{zurek}. 
The factor in brackets is a scattering
rate of a condensate atom on noncondensate atoms - the very process by
which the thermal cloud environment learns the quantum state of the
system.

  Let us estimate the decoherence rate for a set of typical parameters:
$T=1\mu$K, $\omega=50$Hz, and $a=3\dots 5$nm. The thermal velocity is
$v_{\rm T} \approx 10^{-2}$m/s. The volume of the trap can be approximated
by $V=4\pi a_{\rm return}^3/3$, where $a_{\rm return}=\sqrt{2 k_{\rm B}T/m
\omega^2}$ is a return point in a harmonic trap at the energy of $k_{\rm
B}T$. We estimate the decoherence time as $t_{\rm dec} \approx 10^{5} {\rm
sec} / ( N_{\rm E} N^2)$. For $N_{\rm E}=10^0\dots 10^4$ and $N=10\dots
10^7$ it can range from $1000$s down to $10^{-13}$s. For $N=10$ our
(over-)estimate for $t_{\rm dec}$ is large. However, already for $N=1000$
and $N_{\rm E}=10$ (which are still below the Thomas-Fermi regime)  
$t_{\rm dec}$ shrinks down to milliseconds. Given that our $t_{\rm dec}$
is an upper estimate and that big condensates are more interesting as
Schr\"odinger cats, it is clear that for the sake of cat's
longevity, one must go beyond the standard harmonic trap setting.

  From Eqs.(\ref{dr},\ref{NE}) it is obvious that the decoherence rate
depends a lot on temperature and on chemical potential. The two factors
influence strongly both $N_{\rm E}$ and $v_{\rm T}$. Both can be improved
by the following scenario, which is a combination of present day
experimental techniques. In the experiment of Ref.\cite{dip} a narrow
optical dip was superposed at the bottom of a wide magnetic trap. The
parameters of the dip were tuned so that it had just one bound state. The
gap between this single condensate mode and the first excited state was
$1.5\mu$K, which at $T=1\mu$K gives a fugacity of $z=\exp(-1.5)$. 
We need the gap so that we can use the single mode approximation. At low
temperatures, the gap results in a small fugacity, which is convenient 
for calculations.
We propose to prepare a condensate inside a similar combination of a wide
magnetic and a narrow optical trap (or more generally: a wide well plus a
narrow dip with a single bound mode) and then to open the magnetic trap
and let the noncondensed atoms disperse. The aim is to get rid of the
thermal cloud as much as possible. A similar technique was used in the
experiment of Ref.\cite{optical}.

  Let us estimate the ultimate limit for the efficiency of this technique.  
At the typical initial temperature of $1\mu$K the thermal velocity of
atoms is $10^{-2}$m/s. An atom with this velocity can cross a $1\mu$m dip
in $10^{-4}$s. If we wait for, say, $1$s after opening the wide trap, then
all atoms with velocities above $10^{-6}$m/s will disperse away from the
dip.  A thermal velocity of $10^{-6}$m/s corresponds to the temperature of
$10^{-8}\mu$K. As the factor $N_E v_T \sim T^{7/2}$ in Eq.(\ref{dr}), then
already $1$s after opening the wide trap the decoherence rate 
due to non-condensed atoms is reduced
by a factor of $10^{-28}$!

  It is not realistic to expect such a ``cosmological" reduction factor.
The ``dip" which is left after the wide harmonic trap is gone could be, for
example, a superposition of an ideal dip plus a wide shallow well (which
was a negligible perturbation in presence of the wide harmonic trap).  
The well would have a band of width $\Delta E$ of bound states which would
not disperse but preserve their occupation numbers from before the opening
of the wide trap. They would stay in contact with the condensate and continue
to ``monitor'' its quantum state. 
Even if such a truncated environment happens
to be already relatively harmless, there are means to do better than that.

Further reduction of the decoherence rate can be achieved by
``symmetrization" of the environmental state. Perfect symmetrization can be
obtained provided that:

i)   Atoms have two internal states $|A\rangle$ and $|B\rangle$. 

ii)  $|A\rangle$ and $|B\rangle$ feel the same trap potential. 

iii) $A-A$ and $B-B$ scattering lengths are the same. 

iv)  A Hamiltonian has a term which drives coherent transitions
     $A\leftrightarrow B$ with a frequency of $\lambda/\hbar$.

v) $ \Delta E \; \ll \lambda$. \\
The term in (iv) can be realized by driven coherent transitions like 
in the experiments of Refs.\cite{lambda}.

Given an ideal symmetry between $A$ and $B$ (assumptions i-iii), the
eigenmodes of the dilute environment have annihilation operators $S_s \sim
(a_s+b_s)$ and $O_s \sim (a_s-b_s)$, which are symmetric and antisymmetric
respectively. $a_s$ and $b_s$ are annihilation operators for the two
internal states of an atom in the trap eigenstate numbered as ``$s$" with
trap eigenenergy of $\epsilon_s$. $S_s$ has energy $(\epsilon_s-\lambda)$
and $O_s$ has energy $(\epsilon_s+\lambda)$. The symmetric and
antisymmetric $\Delta E$-bands of states can be visualized as two ladders
shifted with respect to each other by $2 \lambda$. In other words, the two
sets of states feel the same, but shifted, trapping potentials. If (v) is
satisfied, then the antisymmetric $O_s$'s are nearly empty since they can
evaporate into symmetric states and then leave the trap. The symmetric
$S_s$'s cannot distinguish between $A$ and $B$ so they do not destroy the
quantum coherence between the Schr\"odinger cat's components. After
symmetrization, $N_{\rm E}$ in Eq.(\ref{dr}) has to be replaced by the
final number of atoms in the antisymmetric states only;

\begin{equation}\label{NEO}
N_{\rm E}^{\rm O}
\;\approx\;
\left\{
\begin{array}{ll}
n_{\lambda} \exp( (\mu-\lambda)/ k_{\rm B} T ),  
& \mbox{for $2 \lambda < \Delta E$} \\
0,  & \mbox{for $2 \lambda > \Delta E$}
\end{array}
\right.
\end{equation}
Here $T$ is the temperature before opening the wide trap. $n_{\lambda}$ is
the number of antisymmetric bound states which remain within the $\Delta
E$-band of symmetric states. Atoms in these antisymmetric bound states
cannot disperse away. For $2\lambda > \Delta E$ this number $n_{\lambda}$
is zero and there is no decoherence from the thermal cloud.

In this perfect symmetrization limit the states 
$|\pm\rangle \equiv (|N,0\rangle\pm|0,N\rangle )/\sqrt{2}$ 
exist within a 
decoherence-free pointer subspace of the Hilbert space, since they have 
degenerate eigenvalues of the interaction hamiltonian $V$ 
\cite{pointer,paolo,lidar}. 
Any state of that subspace can be written as
$\alpha |N,0\rangle + \beta |0,N\rangle$, with $\alpha$ and $\beta$ complex
numbers. If ${\cal{P}}_{[\alpha|N,0\rangle+\beta|0,N\rangle]}$ denotes a 
projector onto that subspace, then 
\begin{equation}
[V, {\cal{P}}_{[\alpha|N,0\rangle+\beta|0,N\rangle]}] = 0, 
\label{conmut}
\end{equation}
which means that any quantum superposition
$\alpha |N,0\rangle + \beta |0,N \rangle$ in the subspace 
is an eigenstate of the system operators in the 
interaction hamiltonian (a perfect pointer state), 
and as such will retain its phase
coherence and last forever. The
interaction Hamiltonian between the condensate and the thermal cloud is a
sum of products of condensate operators and environmental operators. Only
terms with symmetric environmental operators are relevant because
the antisymmetric states are empty.  
The total Hamiltonian is symmetric with respect to $A \leftrightarrow B$
so, to preserve this symmetry, the relevant terms with symmetric
environmental operators also contain symmetric condensate operators. The
argument simplifies a lot for small fugacity where there is only one
leading term with the $N_A+N_B$ condensate operator. The states
$|\pm\rangle$ are its eigenstates with the same eigenvalue $N$. They are
also (almost) degenerate eigenstates of the condensate Hamiltonian build
out of $N_{A,B}$. The coherent transitions $A\leftrightarrow B$ break this
degeneracy of $|\pm\rangle$ but the difference of their eigenfrequencies
is negligible as compared to the usual condensate lifetime of $\sim 10$s.
In the next-to-leading order in fugacity there are symmetric interaction
terms which change the number of condensed atoms. These terms drive the
$|N,0\rangle$ and $|0,N\rangle$ states into slightly ``squeezed-like'' states
$|S,0\rangle$ and $|0,S\rangle$ respectively \cite{moon}. 
There are also terms which exchange $A$ with
$B$. They give each state a small admixture of the opposite component. 
Superpositions of these are still decoherence-free pointer subspaces - 
there are no relevant antisymmetric operators to
destroy their quantum coherence.
When the antisymmetric environmental states begin to be occupied (see 
Eq.(\ref{NEO})), then the commutation relation Eq.(\ref{conmut}) is 
only approximate
and states within the subspace will decohere. To leading order in fugacity and
condensate size, the decoherence rate is given by Eqs. (\ref{dr}) and
(\ref{NEO}).

 This pointer subspace is not perfect -- its existence
is in apparent contradiction with the
finite lifetimes of the condensates which can last for at most $10\dots
20$s. The reason is that the thermal cloud is not the only source of
decoherence. The condensate loses atoms because of Rayleigh scattering,
external heating, and three-body decay. The atoms which escape from the
condensate carry information about its quantum state. They destroy its
quantum coherence. Three body decay, the last process of the three above
is the most important one \cite{dip}. In the experiment of Ref.\cite{dip}
the measured loss rate per atom was $4/$s for $N=10^7$ or around $1$ atom
per $10^{-7}$s. The last rate scales like $N^3$ so already for $N=10^4$
just one atom is lost per second; decoherence time is $1$s. Another
possibility is to increase slightly the dip radius. The loss rate scales
like density squared so an increase in the dip width by a factor of $2$
reduces the loss rate by a factor of $2^6=64$.

  Ambient magnetic fields are yet another source of coherence loss
\cite{private}. The condensed atoms have magnetic moments. If the
magnetic moments of $A$ and $B$ were different the magnetic field would
distinguish between them and would 
introduce an unknown phase into the quantum superposition, 
thus rendering its underlying coherent nature undetectable.
Fortunately the much used $|F,m_F\rangle=|2,1\rangle,|1,-1\rangle$ states
of $^{87}$Rb have the same magnetic moments. For them the magnetic field
is a ``symmetric" environment \cite{private}.

One more source of decoherence is the typical $\approx 1 \%$ difference
between the $A-A$ and $B-B$ scattering lengths which violates the
assumption (iii) above. The Hamiltonian is not perfectly symmetric under
$A\leftrightarrow B$. Even for a perfectly symmetrized environment
symmetric environmental operators couple to not fully symmetric condensate
operators.  This means that for the $1 \%$ difference of scattering lengths
symmetrization can improve decoherence time by at most two orders of
magnitude as compared to the unsymmetrized environment.

 In summary, we outlined a BEC scenario for an experimental realisation of
a decoherence-free Schr\"odinger cat. This scenario has two ingredients:

1) opening of the wide trap followed by an evaporation of the thermal
cloud,

2) and symmetrization of the environment.\\ 
The Schr\"odinger cat is expected to be a quantum superposition of number
eigenstates, $|N,0\rangle$ and $|0,N\rangle$.

More details can be found in Sections II-IV, where we study the
decoherence rate and the idea of symmetrized environment. Finally, Section
V contains discussions.

\section{ Symmetrized Environment }

  We introduce annihilation operators for $|A\rangle$ and $|B\rangle$,

\begin{eqnarray}\label{phiAB}
\phi_{A}(\vec{x})\;=
\;a\;g(\vec{x})\;+\;
\sum_{s}\;a_{s}\;u_{s}(\vec{x})\;, \nonumber\\
\phi_{B}(\vec{x})\;=
\;b\;g(\vec{x})\;+\;
\sum_{s}\;b_{s}\;u_{s}(\vec{x}) \;\;. 
\end{eqnarray}
where $g(\vec{x})$ is the ground state wave-function localized in the dip
with energy $\epsilon_g<0$, and $s$ is an index running over excited
(environmental) states with ortonormal wave-functions $u_{s}(\vec{x})$.
The Hamiltonian of the system is

\begin{equation}\label{H}
H=\int d^3x \left[ 
v(\phi_A^{\dagger}\phi_A)(\phi_B^{\dagger}\phi_B)+ 
\left\{
\frac{u}{2} \phi_A^{\dagger}\phi_A^{\dagger} \phi_A \phi_A +
\nabla\phi_A^{\dagger}\nabla\phi_A+
U(r)\phi_A^{\dagger}\phi_A-
\lambda \phi_A^{\dagger}\phi_B \right\}+                
\left\{ A \leftrightarrow B \right\} \right] .
\end{equation}
Here $u=4 \pi \hbar^2 a_{AA}/m$ ($a_{AA}=a_{BB}$ are the inter-scattering
lengths) and $v=4 \pi \hbar^2 a_{AB}/m$ ($a_{AB}$ is the intra-scattering
length). The immiscibility assumption implies $v>u$. $U(r)$ is the
trap potential and $\lambda$ is the strength of the coherent driving
transitions $A \leftrightarrow B$. Substitution of (\ref{phiAB}) into
(\ref{H}) and subsequent linearization gives a Hamiltonian for the dilute
environment

\begin{equation}\label{HEab}
H_{\rm E}=\sum_s\left[
\epsilon_s(a_s^{\dagger}a_s+b_s^{\dagger}b_s)
-\lambda(a_s^{\dagger}b_s+b_s^{\dagger}a_s)
\right]\;.
\end{equation}
$\epsilon_s$ is the single particle energy of level $s$. We take the
lowest enviromental energy ${\rm min}[\epsilon_s]=0$. A transformation

\begin{equation}
S_s=\frac{a_s+b_s}{\sqrt{2}} \;\;,\;
O_s=\frac{a_s-b_s}{\sqrt{2}} 
\end{equation}
brings $H_E$ to a diagonal form

\begin{equation}\label{HEAB}
H_{\rm E}=\sum_s\left[
(\epsilon_s-\lambda)S_s^{\dagger}S_s+
(\epsilon_s+\lambda)O_s^{\dagger}O_s
\right]\;.
\end{equation}
Symmetric $S$'s and antisymmetric $O$'s form two identical ladders of
states but shifted with respect to each other by $2\lambda$. There is a
gap $(-\lambda -\epsilon_g)>0$ between the lowest $S_s$ state and the
ground state. In equilibrium the state $S_s$'s occupation number is
$n^{\rm S}_s=1/[\exp(\beta(\epsilon_s-\lambda-\epsilon_g))-1]\approx
 z\exp(-\beta(\epsilon_s-\lambda))$, where the fugacity
$z=\exp(-\beta|\epsilon_g|)$ is assumed to be small. The $O_s$'s
occupation number is $n^{\rm O}_s\approx n^{\rm S}_s\exp(-2\beta\lambda)$. 
If $2\lambda\beta \gg 1 $, then $O$'s would be nearly empty. The condensate
two-mode Hamiltonian is

\begin{equation}\label{HC}
H_{\rm C}=\epsilon_g(a^{\dagger}a+b^{\dagger}b)-
    \lambda(a^{\dagger}b+b^{\dagger}a)+               
    \frac{u_{\rm c}}{2}(a^{\dagger}a^{\dagger}aa+b^{\dagger}b^{\dagger}bb)+
    v_{\rm c}(a^{\dagger}b^{\dagger}ab)\;,
\end{equation}
where e.g. $u_{\rm c}=u\int d^3x\;g^4(\vec{x})$ and $v_{\rm c}>u_{\rm c}$. This Hamiltonian
was studied in detail in \cite{lz,steel}. For a purity factor
$\epsilon\equiv(\lambda/(v_{\rm c}-u_{\rm c})N)^N \ll 1$ the lowest energy subspace
contains two macroscopic superpositions

\begin{eqnarray}\label{pure}
&& |+\rangle=
   \frac{1}{\sqrt{2\;N!}}[ (a^{\dagger})^N + (b^{\dagger})^N ] \; |0,0\rangle
   \equiv
   \frac{1}{\sqrt{2}}( |N,0\rangle + |0,N\rangle ) \;\;,\nonumber\\
&& |-\rangle=
   \frac{1}{\sqrt{2\;N!}}[ (a^{\dagger})^N - (b^{\dagger})^N ] \; |0,0\rangle
   \equiv
   \frac{1}{\sqrt{2}}( |N,0\rangle - |0,N\rangle ) \;\;.
\end{eqnarray}
The lower $|+\rangle$ and the higher $|-\rangle$ states are separated by a
small energy gap of $N (u_{\rm c}-v_{\rm c})\epsilon\ln\epsilon$. 
If we just have $\epsilon<1$,
the $|\pm\rangle$ states contain an admixture of intermediate states
$|N-1,1\rangle,\ldots,|1,N-1\rangle$ such that their overlap is $\langle +
| - \rangle =\epsilon$. For $\epsilon\gg 1$ they shrink to

\begin{equation}\label{dirty}
(a^{\dagger} \pm b^{\dagger})^N \; |0,0\rangle \;\;.
\end{equation}
>From now on we assume the pure case, $\epsilon\ll 1$. 

Finally, the interaction Hamiltonian $V$, which contains all
condensate-noncondensate vertices, is a sum of a symmetric

\begin{eqnarray}\label{Vs}
V_{\rm S} =  
&&
\left[(4u+2v)(a^{\dagger}a+b^{\dagger}b)+(2v)(a^{\dagger}b+b^{\dagger}a)\right]
\otimes
\left[\sum_{s_1,s_2}S_{s_1}^{\dagger}S_{s_2}\alpha_{s_1^{\star}s_2}\right]+
\nonumber\\
&& 
\left[ (u+v)(a^{\dagger}+b^{\dagger}) \right] 
\otimes 
\left[ \sum_{s_1s_2s_3}S^{\dagger}_{s_1} S_{s_2}S_{s_3}
       \beta_{s_1^{\star}s_2 s_3} \right]+ {\rm h.c}.+ 
\nonumber\\
&&
\left[(4u+2v)(a^{\dagger}a+b^{\dagger}b)-(2v)(a^{\dagger}b+b^{\dagger}a)\right]
\otimes
\left[\sum_{s_1,s_2}O_{s_1}^{\dagger}O_{s_2}\alpha_{s_1^{\star}s_2}\right]+
\nonumber\\
&&
\left[a^{\dagger}+b^{\dagger}\right]
\otimes
\left[(u-v)\sum_{s_1s_2s_3} S^{\dagger}_{s_1} O_{s_2} O_{s_3}
      \beta_{s_1^{\star}s_2 s_3}+
      (4u)\sum_{s_1s_2s_3}O^{\dagger}_{s_1}O_{s_2} S_{s_3}
      \beta_{s_1^{\star}s_2 s_3}\right]
+ {\rm h.c.}    
\end{eqnarray}
and an antisymmetric part

\begin{eqnarray}\label{Va}
V_{\rm O}&=&
\left[(4u-2v)(a^{\dagger}a-b^{\dagger}b)+
      (2v)(a^{\dagger}b-b^{\dagger}a)\right]
\otimes
\left[\sum_{s_1,s_2}S_{s_1}^{\dagger}O_{s_2}\alpha_{s_1^{\star}s_2}
      + {\rm h.c.}\right]+
\nonumber\\
&&
\left[a^{\dagger}-b^{\dagger} \right]
\otimes
\left[\sum_{s_1s_2s_3}\beta_{s_1^{\star}s_2 s_3}
      ((u+v)O^{\dagger}_{s_1}O_{s_2} O_{s_3}+
      (u-v)O^{\dagger}_{s_1}S_{s_2} S_{s_3}+
      (2u)S^{\dagger}_{s_1} S_{s_2} O_{s_3})\right]+ {\rm h.c.}
\end{eqnarray}
The coefficients are given by integrals

\begin{equation}
\alpha_{s_1^{\star}s_2}=\frac{1}{4}
  \int d^3x\;g^2u_{s_1}^{\star}u_{s_2}\;\;,\;
\beta_{s_1^{\star}s_2^{\star}s_3}=\frac{1}{2\sqrt{2}}
  \int d^3x\;gu_{s_1}^{\star}u_{s_2}^{\star}u_{s_3}\;.
\end{equation}
In $V_{\rm S}$ and $V_{\rm O}$ we neglected vertices ${\rm C}+{\rm
C}\rightarrow {\rm C}+ {\rm NC}$ and ${\rm C}+{\rm C}\rightarrow {\rm NC}+
{\rm NC}$, where ${\rm C}$ is a condensate and ${\rm NC}$ is a
noncondensate particle, and their hermitian conjugates. They are forbiden
by energy conservation due to the gap between the condensate mode and the
lowest environmental state.

  All the terms in $V_{{\rm S},{\rm O}}$ were arranged in the form
[condensate operator]$\otimes$[environment operator]. $V_{\rm S}$ contains
only [c.o.]'s which are symmetric under $a \leftrightarrow b$. They act in
precisely the same way on both components of the superposition and as such
they do not destroy coherence between the macroscopic components. 
To illustrate this let us calculate a commutator of the leading order term 
in $V_S$ 
with the projector onto the subspace of states 
$\alpha|N,0\rangle+\beta|0,N\rangle$,

\begin{equation}
\left[ (4u+2v)(a^{\dagger}a+b^{\dagger}b)
        \otimes
        \sum_{s_1,s_2}S_{s_1}^{\dagger}S_{s_2}\alpha_{s_1^{\star}s_2}
        \;,\;
        {\cal{P}}_{[\alpha|N,0\rangle+\beta|0,N\rangle]}
\right]=0
\end{equation}
This commutator vanishes because states of the form 
$\alpha|N,0\rangle+\beta|0,N\rangle$ are eigenstates of 
$N_{\rm C}=a^{\dagger}a+b^{\dagger}b$. Therefore, this subspace would be a 
pointer subspace \cite{pointer} if this leading term were the only term 
in the interaction Hamiltonian. However, $V_{\rm S}$ has other terms 
that are not simply functions of the total condensate number operator. 
Coherent states of an annihilation operator $a+b$ are exact eigenstates 
of the [c.o.] $a+b$ and approximate eigenstates of $a^{\dagger}+b^{\dagger}$.
These coherent states, however, are combinations of (\ref{dirty}) and as 
such they are not in the lowest energy subspace of $H_{\rm C}$. What is more,
the matrix elements of $a+b$ are of the order $O(\sqrt{N})$ which is 
negligible as compared to the matrix elements of the number operator. If we
 project $a+b$ and its h.c. on the subspace (\ref{pure}), then their 
approximate eigenstates for large $|z|$ are coherent states
 $\alpha|z,0\rangle+\beta|0,z\rangle$. The decoherence effects of $a+b$ and
 $N_{\rm C}$ put together lead to a superposition of macroscopic 
``squeezed-like'' states $\alpha|S,0\rangle+\beta|0,S\rangle$. This result 
is by now well established for a single component condensate \cite{moon}.
 Finally, the [c.o.] $a^{\dagger}b+b^{\dagger}a$ drives 
the state out of the pointer
subspace (\ref{pure}). Its effect is supressed by the purity condition 
$\epsilon\ll 1$ and is also negligible as compared to the direct effect of 
the $\lambda$-term in $H_C$. Similar comment applies to the ``out of the 
subspace" action of $a^{\dagger}+b^{\dagger}$. Therefore, the effect of all 
these terms in $V_{\rm S}$ imply that the subspace spanned by $|N,0\rangle$ 
and $|0,N\rangle$ is no longer an exact decoherence-free pointer subspace,
 i.e. 
$[V_{\rm S},  {\cal{P}}_{[\alpha|N,0\rangle+\beta|0,N\rangle]}   ] \neq 0$. 
The correct pointer subspace would be one spanned by those ``squeezed-like'' 
states. However, to leading order in fugacity and in the condensate size,
 $N$, the dominant terms are those depending on $N_{\rm C}$; the subspace 
of $\alpha|N,0\rangle+\beta|0,N\rangle$ is (approximately) the exact 
decoherence-free pointer subspace.

  The antisymmetric [c.o.]'s in $V_{\rm O}$ act in opposite way on both
components; they destroy their quantum coherence. To illustrate this let us 
calculate a commutator of the leading term in $V_{\rm O}$ with the projector
operator onto the subspace $\alpha|N,0\rangle+\beta|0,N\rangle$,

\begin{eqnarray}
&&
\left[
(4u-2v)(a^{\dagger}a-b^{\dagger}b)
\otimes
\left(\sum_{s_1,s_2}S_{s_1}^{\dagger}O_{s_2}\alpha_{s_1^{\star}s_2}
      + {\rm h.c.}\right)
\;,\;
{\cal{P}}_{[\alpha|N,0\rangle+\beta|0,N\rangle]}
\right]= \nonumber \\
&&
2N(4u-2v)
\left(
\alpha\beta^{\star} |N,0\rangle\langle 0,N|-
\alpha^{\star}\beta |0,N\rangle\langle N,0|
\right)
\otimes
\left(\sum_{s_1,s_2}S_{s_1}^{\dagger}O_{s_2}\alpha_{s_1^{\star}s_2}
      + {\rm h.c.}\right)
\end{eqnarray}
The commutator is $O(N)$. The occupation numbers of $O$'s are supressed by a 
Boltzmann factor $n^{\rm S}\exp(-2\lambda\beta)$; for $2\lambda\beta\gg 1$
 the $O$'s are unoccupied. As such they cannot scatter into condensate 
particles. It is also impossible to scatter $S$'s into $O$'s thanks to energy
 conservation. The unoccupied $O$'s are irrelevant for decoherence and as
 such can be neglected in $V$ and in the above commutator. This effectively 
sets the dangerous $V_{\rm O}$ to zero and leaves us only with 
$V_{\rm S}\neq 0$. In the absence of $O$'s the 
states $\alpha|N,0\rangle+\beta|0,N\rangle$ are a pointer subspace (or a 
decoherence-free subspace). The symmetrized environment of just $S$'s 
defines the quantum states of the components but it does not destroy their 
mutual coherence.

\section{ Master Equation }

  Symmetrized environment is a robust idea whose validity does not depend
on detailed calculations. Nevertheless, for the sake of illustration we
derived (by a perturbative expansion in $V$) an approximate Bloch-Lindblad
form master equation for the reduced density matrix $\rho(t)$ of the
condensate modes. The calculations are long but rather straightforward,
their details can be found in Section IV. Here we just give the final
result.

\begin{equation}
\dot{\rho}=\frac{i}{\hbar} [\rho,H_C^{\rm ren}] + \dot{\rho}_{{\rm S}} + 
\dot{\rho}_{\rm O} \;\;.
\end{equation}
$H_C^{\rm ren}$ is a renormalized condensate Hamiltonian, 

\begin{equation}
H_C^{\rm ren} = H_C + \langle H_C \rangle +
                      c_1 \Delta_1^{\dagger} \Delta_1 + 
                      c_2 \Delta_2^{\dagger} \Delta_2 +  
                      c_3 \Delta_3^{\dagger} \Delta_3 +  
                      c_4 \Delta_4^{\dagger} \Delta_4 \;\;.
\end{equation}
$\langle\dots\rangle$ means an average over an initial environment thermal
density matrix and $\dot{\rho}_{{\rm S}}$ and $\dot{\rho}_{{\rm O}}$ are
contributions from $V_{\rm S}$ and $V_{\rm O}$ respectively. They read

\begin{eqnarray}
\dot{\rho}_{\rm S} &=& d_1 D[{\Delta}_1] \rho  + 
   		     d_2 D[{\Delta}_2] \rho \;\;,\\
\dot{\rho}_{\rm O} &=& d_3 D[{\Delta}_3] \rho  +
                    {\tilde d}_3 D[{\Delta}_3^{\dagger}] \rho +
		     d_4 D[{\Delta}_4] \rho
\end{eqnarray}
with $D[\Delta] \rho \equiv \{ \Delta^{\dagger} \Delta, \rho \} - 2 \Delta
\rho \Delta^{\dagger}$ the Lindblad operator. The operators introduced in
this expression are defined as

\begin{eqnarray}
\Delta_1 &=& (4u + 2v) (a^{\dagger} a + b^{\dagger} b) + (2v) (a^{\dagger} b + 
             b^{\dagger} a) \;\;,\\
\Delta_2 &=& (u+v) (a^{\dagger}+ b^{\dagger}) \;\;,\\
\Delta_3 &=& (4u - 2v) (a^{\dagger} a - b^{\dagger} b ) + (2v) (a^{\dagger} b-
             b^{\dagger} a) \;\;,\\
\Delta_4 &=& a^{\dagger} - b^{\dagger} \;\;.
\end{eqnarray}
The coefficients $c_i$ and $d_i$ are proportional to fugacity $z$. What is
more $c_{3,4}/c_{1,2}\sim e^{-2 \lambda\beta }$ and $d_{3,4}/d_{1,2}\sim
e^{-2 \lambda\beta }$. As anticipated, the antisymmetric contributions are
supressed by a Boltzmann factor due to $O$'s. This approximate master
equation confirms our heuristic arguments.

  Any asymmetry between $A$ and $B$ is a source of decoherence.  If, for
example, the $A-A$ and $B-B$ scattering lengths were different,
$u_{AA}=u+\delta u$ and $u_{BB}-\delta u$, then $V_{\rm O}$ would acquire
extra terms $\sim\; \delta u$. They would show up in $\dot{\rho}_{\rm O}$
with coefficients $d\sim z \; u\delta u$. The ratio of these to the
``symmetric'' coefficients $d_{1,2}$ is $\delta u/u$ (which is
approximately $0.03$ for $|F=1;m_F=-1\rangle$ and $|F=2;m_F=1\rangle$
states of $^{87}$Rb ) as compared to $d_{3,4}/d_{1,2}\sim
\exp(-2\lambda\beta)$. If $\delta u/u$ is less than the Boltzmann factor,
then this asymmetry is not a leading source of decoherence.

\section{ Derivation of the Master Equation }

 We derive a master equation by perturbative expansion in $V$ or in the
coupling constants $u,v$ which we regard to be of the same order.  We
assume the initial density matrix at $t=0$ to be a product
$\tilde{\rho}(0)=\rho(0)\otimes\rho_E$ of the system (condensate) and
environment (noncondensate) density matrices. To take a more accurate
starting point we make a rearrangement $V \rightarrow V-\langle V \rangle$
and $H_{\rm C} \rightarrow H_{\rm C}+ \langle V \rangle$ where
$\langle\dots\rangle={\rm Tr}_{\rm E}[\dots\rho_{\rm E}]$ is a trace over
the environment thermal density matrix at the initial time. The new
$H_{\rm C}$ differs from the old one by a renormalization

\begin{eqnarray}\label{eff} && \epsilon_g \rightarrow
   \epsilon_g^{\rm eff}=\epsilon_g+(4u+2v)\sum_s n^{\rm S}_s 
\alpha_{s^{\star}s}
   \;\;,
\nonumber\\
&& \lambda \rightarrow
   \lambda^{\rm eff}=\lambda-(2v)\sum_s n^{\rm S}_s \alpha_{s^{\star}s} \;.
\end{eqnarray}
We compute $\dot{\tilde{\rho}}(t)$ up to second order in the perturbation
Hamiltonian $V = \sum_i \Delta_i \otimes E_i$, with $\Delta_i$ an operator
for the condensate and $E_i$ one for the environment. The master equation
for $\rho(t)$ alone is obtained by tracing $\dot{\tilde{\rho}}(t)$ over
the non-condensed modes. The result is \cite{wm}

\begin{equation}
{\dot \rho}(t) = \frac{i}{\hbar} [\rho, H_{\rm C}] 
- \frac{1}{2 \hbar^2} \sum_{ij} \int_0^t d\tau g^{\rm sym}_{ij}(\tau) 
[\Delta_i(0),[\Delta_j(-\tau),\rho]]
- \frac{1}{2 \hbar^2} \sum_{ij} \int_0^t d\tau g^{\rm asym}_{ij}(\tau) 
[\Delta_i(0),\{\Delta_j(-\tau),\rho \}]\;,
\label{master}
\end{equation}
where

\begin{eqnarray}
g^{\rm sym}_{ij}(\tau) &=& \langle \{ E_i(\tau), E_j(0) \} \rangle \;\;,\\
g^{\rm asym}_{ij}(\tau) &=& \langle [E_i(\tau), E_j(0) ] \rangle \;\;.
\end{eqnarray}
We have used that after the rearrangement $V \rightarrow V-\langle V
\rangle$ and $H_{\rm C} \rightarrow H_{\rm C} + \langle V \rangle$, there
are no linear terms in the perturbative expansion because $\langle V
\rangle=0$. We first start to study the contribution of the symmetric
terms $V_{\rm S}$ in the interaction Hamiltonian, and later we shall deal
with the antisymmetric ones $V_{\rm O}$.  It is easy to show that the are
no cross terms S-O in the calculation of the master equation.

\subsection{Free Hamiltonian}

The next step in the calculation is to solve the dynamics of the free
condensate Hamiltonian

\begin{equation}
H_{\rm C}=(u_{\rm c}- v_{\rm c}) J_z^2 + \lambda^{\rm eff} J_x \;\;,
\end{equation}
where we introduced angular momentum operators

\begin{eqnarray}
J_x &=& \frac{1}{2} ( a^{\dagger} b + b^{\dagger} a) \;\;,\\
J_y &=& \frac{i}{2} (b^{\dagger} a - a^{\dagger} b)  \;\;,\\
J_z &=& \frac{1}{2} (a^{\dagger} a - b^{\dagger} b ), 
\end{eqnarray}
and discarded constant terms proportional to the total number of particles
$N=a^{\dagger} a + b^{\dagger} b$ \cite{steel}. The Heisenberg equations
of motion for the condensate operators $a(t)$ and $b(t)$ are

\begin{eqnarray}
&& i \hbar \dot{a}=\epsilon_g^{\rm eff} a - \lambda^{\rm eff} b 
            + (u_{\rm c} a^{\dagger}a + v_{\rm c} b^{\dagger}b)a
\;,\nonumber\\
&& i \hbar \dot{b}=\epsilon_g^{\rm eff} b - \lambda^{\rm eff} a 
            + (u_{\rm c} b^{\dagger}b + v_{\rm c} a^{\dagger}a)b \;.
\end{eqnarray}
These equations cannot be solved exactly for $\lambda^{\rm eff}\neq 0$,
when numbers of $a$'s and $b$'s are not conserved independently due to the
coherent transfer of particles from one state to the other. In their
mean-field version, these equations correspond to the well-known
macroscopic self-trapping equation, which have the feature that below a
critical ``purity'' the oscillations between states $A$ and $B$ are not
complete; for $\epsilon \ll 1$ the systems is self-locked in either of
these states \cite{lock}.  In this limit we can set $\lambda^{\rm eff}$ to
zero, and for the immiscible case $v_{\rm c} >u_{\rm c}$ the ground state
of $H_C$ corresponds to maximum eigenstates of $J_z^2$, i.e.  pure cat
states (\ref{pure}). Also in this case the mean-field solutions are

\begin{equation}
a(t)=a(0)e^{-\frac{i t}{\hbar} [\epsilon_{\rm g}^{\rm eff}+N(u_{\rm c}+
v_{\rm c})/2]} \;\;,
b(t)=b(0)e^{-\frac{i t}{\hbar} [\epsilon_{\rm g}^{\rm eff}+N(u_{\rm c}+
v_{\rm c})/2]} \;\;.
\end{equation}
We believe that these solutions are qualitatively correct as long as the
purity factor $\epsilon\ll 1$.

\subsection{Symmetric Interaction Hamiltonian}

The symmetric Hamiltonian can be written as
\begin{equation}
V_{\rm S}= \Delta_1 E_1 + {\tilde \Delta}_1 {\tilde E}_1 + 
\left[ \Delta_2 E_2 + {\tilde \Delta_2} {\tilde E}_2 + 
{\rm h.c.} \right] \;\;,
\end{equation} 
where we have defined condensate operators
\begin{eqnarray}
\Delta_1 &=& (4u + 2v) (a^{\dagger} a + b^{\dagger} b) + (2v) 
(a^{\dagger} b + b^{\dagger} a) \;\;,\\
{\tilde \Delta}_1 &=& (4u + 2v) (a^{\dagger} a + b^{\dagger} b) - (2v) 
(a^{\dagger} b + b^{\dagger} a) \;\;,\\
\Delta_2 &=& (u+v) (a^{\dagger}+ b^{\dagger}) \;\;,\\
{\tilde \Delta}_2 &=& (u+v)(a^{\dagger}+ b^{\dagger}) \;\;, 
\end{eqnarray}
and the corresponding non-condensate operators
\begin{eqnarray}
E_1 &=& \sum_{s_1,s_2} \alpha_{s_1^*,s_2} S_{s_1}^{\dagger} S_{s_1} - 
        \sum_s |\alpha_{s^*,s}|^2 S_s^{\dagger} S_s \;\;,\\
{\tilde E}_1 &=& \sum_{s_1,s_2} \alpha_{s_1^*,s_2} 
O_{s_1}^{\dagger} O_{s_1} - \sum_s |\alpha_{s^*,s}|^2 O_s^{\dagger} O_s \;\;,\\
E_2 &=& \sum_{s_1,s_2,s_3} \beta_{s_1^*,s_2,s_3} 
(S_{s_1}^{\dagger} S_{s_2} S_{s_3} + O_{s_1}^{\dagger} S_{s_2} O_{s_3}) \;\;,\\
{\tilde E}_2 &=&  \sum_{s_1,s_2,s_3} \beta_{s_1^*,s_2,s_3} 
(O_{s_1}^{\dagger} O_{s_2} S_{s_3} + S_{s_1}^{\dagger} O_{s_2} O_{s_3}) \;\;.
\end{eqnarray}
It is clear that the terms $\Delta_i$ and ${\tilde \Delta}_i$ and their
corresponding environment operators have the same structure, and therefore
will give the same qualitatively contribution to the master equation. In
the following we shall keep only the $\Delta_i$ terms.

The expectation values of multiple-points noncondensate operators  
are written in terms of the two-point functions
\begin{eqnarray}
&&
\langle S_{s_1}^{\dagger}(t_1) S_{s_2}(t_2) \rangle=
\delta_{s_1s_2}
n^{\rm S}_{s_1}
e^{ \frac{i}{\hbar} (\epsilon_{s_1}-\lambda)(t_1-t_2)}
e^{-\gamma^{\rm S}_{s_1}|t_1-t_2|}
\;\;,\nonumber\\
&&
\langle O_{s_1}^{\dagger}(t_1) O_{s_2}(t_2) \rangle=
\delta_{s_1s_2}
n^{\rm O}_{s_1}
e^{ \frac{i}{\hbar} (\epsilon_{s_1}+\lambda)(t_1-t_2)}
e^{-\gamma^{\rm O}_{s_1}|t_1-t_2|}
\;\;,\nonumber\\
&&
\langle S_{s_1}^{\dagger}(t_1) {\rm O}_{s_2}(t_2) \rangle=0
\nonumber\\
\end{eqnarray}
via Wick's theorem. Here $n_s^{{\rm S},{\rm O}}=[z e^{\beta(\epsilon_s \pm
\lambda)}-1]^{-1}$ are Bose occupation numbers, and the $\gamma_s$'s are
inverse finite lifetimes of the environmental states. We also expand these
expectation values to leading order in fugacity $z$. The kernels are

\begin{eqnarray}
\langle E_1(\tau) E_1(0) \rangle &=&
\langle E_1(0) E_1(\tau) \rangle^* = 
\sum_{s_1,s_2} |\alpha_{s_1^*,s_2}|^2  n^{\rm S}_{s_1} 
e^{ \frac{i \tau}{\hbar} (\epsilon_{s_1}-\epsilon_{s_2})}
e^{-\tau(\gamma^{\rm S}_{s_1} + \gamma^{\rm S}_{s_2})}  + O(z^2) \;,\\
\langle E_2(\tau) E_2^{\dagger}(0) &=& 
\langle E_2(0) E_2^{\dagger}(\tau) \rangle^* =
\sum_{s_1,s_2,s_3} |\beta_{s_1^*,s_2,s_3}|^2 
e^{\frac{i\tau}{\hbar} (\epsilon_{s_1}-\epsilon_{s_2}-\epsilon_{s_3}+\lambda)}
\left[
2 n^{\rm S}_{s_1} e^{ -\tau(\gamma^{\rm S}_{s_1}+
\gamma^{\rm S}_{s_2}+\gamma^{\rm S}_{s_3})} +
\right. \nonumber \\
&& ~~~~~~~~~~~~~~~~~~~~~~~~~~~~~~~~~~~ \left. 
  n^{\rm O}_{s_1} e^{ -\tau(\gamma^{\rm O}_{s_1}+
\gamma^{\rm S}_{s_2}+\gamma^{\rm O}_{s_3})}
\right] + O(z^2) \;\;,\\
\langle E_2^{\dagger}(\tau)  E_2 (0) \rangle
&=& \langle E_2^{\dagger}(0)  E_2 (\tau) \rangle = O(z^2) \;\;.
\end{eqnarray}
Given the kernels and the free evolution of condensate operators, we have
to perform the time integrals in (\ref{master}). We calculate the master
equation for late times $t \gg \gamma_s^{-1}$. Hence the exponential decay
of the propagators dominates their behavior, and we can apply the
Markovian approximation replacing the upper limit in the time integration
by infinity. This calculation results in two different contributions, one
being a renormalization of the free Hamiltonian

\begin{equation}
\delta H_{\rm C} = c_1 \Delta_1^{\dagger} \Delta_1 + c_2 \Delta_2^{\dagger} 
\Delta_2 \;\;,
\end{equation}
where
\begin{eqnarray}
c_1 &=& \frac{1}{\hbar^2} 
\sum_{s_1,s_2} |\alpha_{s_1^*,s_2}|^2 n^{\rm S}_{s_1} 
\frac{\hbar^{-1} (\epsilon_{s_1}-\epsilon_{s_2})}
{(\gamma^{\rm S}_{s_1}+\gamma^{\rm S}_{s_2})^2+(\hbar^{-1}
(\epsilon_{s_1}-\epsilon_{s_2}))^2} \;\;,\\
c_2 &=& \frac{1}{\hbar^2} \sum_{s_1,s_2,s_3} |\beta_{s_1^*,s_2,s_3}|^2
\left[ 
2 n^{\rm S}_{s_1} \frac{
\hbar^{-1} (\epsilon_{s_1}-\epsilon_{s_2}-\epsilon_{s_3}+ \lambda+
\epsilon^{\rm eff}_{\rm g}
+N(u_{\rm c}+v_{\rm c})/2) }
{(\gamma^{\rm S}_{s_1}+\gamma^{\rm S}_{s_2}+\gamma^{\rm S}_{s_3})^2 + 
(\hbar^{-1}(\epsilon_{s_1}-\epsilon_{s_2}-\epsilon_{s_3}+ \lambda+
\epsilon_{\rm g}^{\rm eff}+N(u_{\rm c}+v_{\rm c})/2))^2 }
+  \right . \nonumber \\
&& ~~~~~~~~~~~~~~~~~~~~~~~~~~~~~ \left. n^{\rm O}_{s_1} \frac{
\hbar^{-1}(\epsilon_{s_1}-\epsilon_{s_2}-\epsilon_{s_3}+ \lambda-
\epsilon^{\rm eff}_{\rm g}
-N(u_{\rm c}+v_{\rm c})/2) }
{(\gamma^{\rm O}_{s_1}+\gamma^{\rm S}_{s_2}+\gamma^{\rm O}_{s_3})^2 + 
(\hbar^{-1}(\epsilon_{s_1}-\epsilon_{s_2}-\epsilon_{s_3}+ \lambda-
\epsilon^{\rm eff}_{\rm g}
-N(u_{\rm c}+v_{\rm c})/2))^2 }
\right] \;.
\end{eqnarray}
The other contribution has the Lindblad form 
$D[\Delta]\rho \equiv \{ \Delta^{\dagger} \Delta,\rho \} - 2 \Delta \rho
\Delta^{\dagger}$ and reads
\begin{equation}
{\dot \rho}_{\rm S}= d_1 D[\Delta_1] \rho + d_2 D[\Delta_2] \rho \;\;,
\end{equation}
where
\begin{eqnarray}
d_1 &=& -\frac{1}{\hbar^2} \sum_{s_1,s_2} 
|\alpha_{s_1^*,s_2}|^2 n^{\rm S}_{s_1}
\frac{\gamma^{\rm S}_{s_1} + \gamma^{\rm S}_{s_2}}{
(\gamma^{\rm S}_{s_1} + \gamma^{\rm S}_{s_2})^2 + 
(\hbar^{-1}(\epsilon_{s_1} - \epsilon_{s_2}))^2} \;\;,\\
d_2 &=& -\frac{1}{\hbar^2} \sum_{s_1,s_2,s_3} |\beta_{s_1^*,s_2,s_3}|^2
\left[
2 n^{\rm S}_{s_1} \frac{
\gamma^{\rm S}_{s_1}+\gamma^{\rm S}_{s_2}+\gamma^{\rm S}_{s_3}}
{(\gamma^{\rm S}_{s_1}+\gamma^{\rm S}_{s_2}+\gamma^{\rm S}_{s_3})^2 + 
(\hbar^{-1}(\epsilon_{s_1}-\epsilon_{s_2}-\epsilon_{s_3}+ \lambda+
\epsilon^{\rm eff}_{\rm g}+N(u_{\rm c}+v_{\rm c})/2))^2 }
+  \right . \nonumber \\
&& ~~~~~~~~~~~~~~~~~~~~~~~~~~~~~ \left. n^{\rm O}_{s_1} \frac{
\gamma^{\rm S}_{s_1}+\gamma^{\rm S}_{s_2}+\gamma^{\rm S}_{s_3}}
{(\gamma^{\rm O}_{s_1}+\gamma^{\rm S}_{s_2}+\gamma^{\rm O}_{s_3})^2 + 
(\hbar^{-1}(\epsilon_{s_1}-\epsilon_{s_2}-\epsilon_{s_3}+ \lambda+
\epsilon^{\rm eff}_{\rm g}
+N(u_{\rm c}+v_{\rm c})/2))^2 }
\right]  \;.
\end{eqnarray}

\subsection{Antisymmetric Interaction Hamiltonian}

The antisymmetric Hamiltonian is
\begin{equation}
V_{\rm O}= \Delta_3 E_3 + \Delta_4 E_4 + {\rm h.c.} \;\;,
\end{equation}
where now
\begin{eqnarray}
\Delta_3 &=& (4u-2v) (a^{\dagger} a - b^{\dagger} b) + (2v)
(a^{\dagger} b - b^{\dagger} a) \;\;,\\
\Delta_4 &=& a^{\dagger} - b^{\dagger} \;\;,
\end{eqnarray}
and the corresponding environment operators
\begin{eqnarray}
E_3 &=& \sum_{s_1,s_2} \alpha_{s_1^*,s_2} O_{s_1}^{\dagger} S_{s_2} \;\;,\\
E_4 &=& \sum_{s_1,s_2,s_3} \beta_{s_1^*,s_2,s_3} \left[
(u+v) O^{\dagger}_{s_1} O_{s_2} O_{s_3} - 
(u-v) O^{\dagger}_{s_1} S_{s_2} S_{s_3} + 
2u S^{\dagger}_{s_1} S_{s_2} O_{s_3}
\right] \;\;.
\end{eqnarray}

The kernels are 
\begin{eqnarray}
\langle E_3(\tau) E_3^{\dagger}(0) \rangle &=&
\langle E_3(0) E_3^{\dagger}(\tau) \rangle^* = 
\sum_{s_1,s_2} |\alpha_{s_1^*,s_2}|^2 n^{\rm O}_{s_1} 
e^{\frac{i \tau}{\hbar} (\epsilon_{s_1}-\epsilon_{s_2} + 2 \lambda)}
e^{-\tau (\gamma^{\rm O}_{s_1} + \gamma^{\rm S}_{s_2})} + O(z^2) \;,\\
\langle E_3^{\dagger}(\tau) E_3(0) \rangle &=&
\langle E_3^{\dagger}(0) E_3(\tau) \rangle^* = 
\sum_{s_1,s_2} |\alpha_{s_1^*,s_2}|^2 n^{\rm O}_{s_1} 
e^{\frac{i \tau}{\hbar} (\epsilon_{s_1}-\epsilon_{s_2} - 2 \lambda)}
e^{-\tau (\gamma^{\rm S}_{s_1} + \gamma^{\rm O}_{s_2})} + O(z^2) \;,\\
\langle E_3^{\dagger}(\tau) E_3(0) \rangle &=& 
\langle E_3^{\dagger}(0) E_3(\tau) \rangle = O(z^2) \;,\\
\langle E_4(\tau) E^{\dagger}_4(0) \rangle &=&
\langle E_4(0) E^{\dagger}_4(\tau) \rangle^*=
\sum_{s_1,s_2,s_3} |\beta_{s^*_1,s_2,s_3}|^2 \left[
2 (u+v)^2 n^{\rm O}_{s_1} e^{\frac{i \tau}{\hbar}
(\epsilon_{s_1}-\epsilon_{s_2}-\epsilon_{s_3}
-\lambda) -\tau (\gamma^{\rm O}_{s_1}+\gamma^{\rm O}_{s_2}+
\gamma^{\rm O}_{s_3})} +
\right. \nonumber \\ 
&& ~~~~~~~~~~~~~~~~~~~~~~~~~~~~~~~~~~~~~~~~~~~~~~~~~~~~~~
\left. 2 (u-v)^2 n^{\rm O}_{s_1} e^{\frac{i \tau}{\hbar} 
(\epsilon_{s_1}-\epsilon_{s_2}-\epsilon_{s_3} +  3\lambda) -
\tau (\gamma^{\rm O}_{s_1}+\gamma^{\rm O}_{s_2}+\gamma^{\rm O}_{s_3})}
\right] \;,\\
\langle E^{\dagger}_4(0) E_4(\tau) \rangle &=& 
\langle E^{\dagger}_4(\tau) E_4(0) \rangle = O(z^2) \;.
\end{eqnarray}
The renormalization terms are
\begin{equation}
\delta H_{\rm C} = c_3 \Delta^{\dagger}_3 \Delta_3 + 
c_4 \Delta^{\dagger}_4 \Delta_4
\end{equation}
with
\begin{eqnarray}
c_3 &=& \frac{1}{\hbar^2} \sum_{s_1,s_2} |\alpha_{s_1^*,s_2}|^2
\left[
n^{\rm O}_{s_1} \frac{\hbar^{-1}(\epsilon_{s_1}-\epsilon_{s_2}-2 \lambda)}{
(\gamma^{\rm S}_{s_1}+\gamma^{\rm O}_{s_2})^2 + 
(\hbar^{-1}(\epsilon_{s_1}-\epsilon_{s_2}-2 \lambda ))^2}
- 
n^{\rm O}_{s_1} \frac{\hbar^{-1}(\epsilon_{s_1}-\epsilon_{s_2}+2 \lambda)}{
(\gamma^{\rm O}_{s_1}+\gamma^{\rm S}_{s_2})^2 + 
(\hbar^{-1}(\epsilon_{s_1}-\epsilon_{s_2}+2 \lambda) )^2} 
\right] \;,\\
c_4 &=& \frac{1}{\hbar^2} \sum_{s_1,s_2,s_3} |\beta_{s_1^*,s_2,s_3}|^2
\left[
2 (u+v)^2 n^{\rm O}_{s_1}  
\frac{\hbar^{-1}(\epsilon_{s_1}-\epsilon_{s_2}-\epsilon_{s_3}-\lambda+
\epsilon^{\rm eff}_{\rm g}
+N(u_{\rm c}+v_{\rm c})/2)  }{
(\gamma^{\rm O}_{s_1}+\gamma^{\rm O}_{s_2}+\gamma^{\rm O}_{s_3})^2 + 
(\hbar^{-1}(\epsilon_{s_1}-\epsilon_{s_2}-\epsilon_{s_3}-
\lambda+\epsilon^{\rm eff}_{\rm g}
+N(u_{\rm c}+v_{\rm c})/2 ))^2} 
+ \right. \nonumber \\
&& \left. ~~~~~~~~~~~~~~~~~~
2 (u-v)^2 n^{\rm O}_{s_1} 
\frac{\hbar^{-1}(\epsilon_{s_1}-\epsilon_{s_2}-\epsilon_{s_3}+
\lambda-\epsilon^{\rm eff}_{\rm g}
-N(u_{\rm c}+v_{\rm c})/2)  }{
(\gamma^{\rm O}_{s_1}+\gamma^{\rm S}_{s_2}+\gamma^{\rm S}_{s_3})^2 + 
(\hbar^{-1}(\epsilon_{s_1}-\epsilon_{s_2}-\epsilon_{s_3}+
3\lambda-\epsilon^{\rm eff}_{\rm g}
-N(u_{\rm c}+v_{\rm c})/2 ))^2} 
\right] \;.
\end{eqnarray}
Finally, the Lindblad part is
\begin{equation}
{\dot \rho}_{\rm O} = d_3 D[\Delta_3] \rho +
 {\tilde d}_3 D[\Delta^{\dagger}_3] \rho
+ d_4 D[\Delta_4] \rho \;\;,
\label{masterodd}
\end{equation}
where
\begin{eqnarray}
d_3 &=& - \frac{1}{\hbar^2} \sum_{s_1,s_2} |\alpha_{s_1^*,s_2}|^2
n^{\rm O}_{s_1} \frac{\gamma^{\rm S}_{s_1}+\gamma^{\rm O}_{s_2}}{
(\gamma^{\rm S}_{s_1}+\gamma^{\rm O}_{s_2} )^2 + 
(\hbar^{-1}(\epsilon_{s_1}-\epsilon_{s_2}-2 \lambda ))^2} \;\;,\\
{\tilde d}_3 & =& - \frac{1}{\hbar^2} \sum_{s_1,s_2} |\alpha_{s_1^*,s_2}|^2
n^{\rm O}_{s_1} \frac{\gamma^{\rm O}_{s_1}+\gamma^{\rm S}_{s_2}}{
(\gamma^{\rm S}_{s_1}+\gamma^{\rm O}_{s_2} )^2 + 
(\hbar^{-1}(\epsilon_{s_1}-\epsilon_{s_2}+2 \lambda ))^2} \;\;,\\
d_4 &=& -\frac{1}{\hbar^2} \sum_{s_1,s_2,s_3} 
|\beta_{s_1^*,s_2,s_3}|^2 \left[
2 (u+v)^2 n^{\rm O}_{s_1}  
\frac{\gamma^{\rm O}_{s_1}+\gamma^{\rm O}_{s_2}+\gamma^{\rm O}_{s_3} }{
(\gamma^{\rm O}_{s_1}+\gamma^{\rm O}_{s_2}+\gamma^{\rm O}_{s_3})^2 + 
(\hbar^{-1}(\epsilon_{s_1}-\epsilon_{s_2}-\epsilon_{s_3}-
\lambda+\epsilon^{\rm eff}_{\rm g}
+N(u_{\rm c}+v_{\rm c})/2 ))^2} 
+ \right. \nonumber \\
&& \left. ~~~~~~~~~~~~~~~~~~~~~~~~~~~~~~~~~
2 (u-v)^2 n^{\rm O}_{s_1} 
\frac{\gamma^{\rm O}_{s_1}+\gamma^{\rm S}_{s_2}+\gamma^{\rm S}_{s_3} }{
(\gamma^{\rm O}_{s_1}+\gamma^{\rm S}_{s_2}+\gamma^{\rm S}_{s_3})^2 + 
(\hbar^{-1} (\epsilon_{s_1}-\epsilon_{s_2}-\epsilon_{s_3}+
3\lambda+\epsilon^{\rm eff}_{\rm g}
+N(u_{\rm c}+v_{\rm c})/2 ))^2} 
\right]  \;.
\end{eqnarray}

\subsection{ Estimate of the Decoherence Rate }

To estimate the decoherence rate for our BEC quantum superposition, let us
consider one of the terms in (\ref{masterodd}), namely $\dot{\rho} = 16
v^2 d_3 [J_z,[J_z,\rho]]$, which arises form the $\Delta_3$ term.  This
term corresponds to elastic two-body collisions between condensate and
non-condensate atoms, and induces phase decoherence. To this end we first
make an estimation for the coefficient $\alpha_{s^*_1,s_2}$ that enters in
$d_3$. We assume that we have a big harmonic trap plus the dip located in
its center. We take as the wave function for the ground state a gaussian
$g({\vec x})= (m \omega_{\rm dip}/(\pi \hbar))^{3/2} \exp(- m \omega_{\rm
dip} r^2/\hbar)$, where $\omega_{\rm dip}$ is the frequency of the optical
trapping potential in the dip. In principle, we should use thermal wave
packets for the wave functions of the non-condensed particles
\cite{gardiner}.  However, due to the strong localization of the
condensate in the dip, we can approximate the non-condensed particles as
plane waves. Then

\begin{equation}
\alpha_{s^*_1,s_2} = \frac{1}{4} \int d^3x g^2(\vec x) 
\frac{e^{i {\vec k}_1 \cdot {\vec x}}}{\sqrt{V}} 
\frac{e^{-i {\vec k}_2 \cdot {\vec x}}}{\sqrt{V}} =
\frac{1}{4 V} \exp\left( 
- \frac{\hbar}{4 m \omega_{\rm dip}} |{\vec k}_1 - {\vec k}_2|^2 \right)
\end{equation}
On the other hand, the decay rate $\gamma$ is much smaller than $w$.  
This lets us replace the last factor in $d_3$ by $ \pi \delta(\hbar^{-1}
(\epsilon_{s_1} - \epsilon_{s_2}))$, where we have used the identity
$\lim_{\gamma \rightarrow 0} \frac{\gamma}{\gamma^2 + x^2} = \pi
\delta(x)$. Finally, replacing the sum over momenta by the continuum
expression $\sum_{\vec k} \rightarrow V \int d^3k= V \int d\Omega dk k^2
$, and using the free dispersion relation $\epsilon_{\vec k} = \hbar^2
k^2/ 2m$, we find

\begin{equation}
d_3=-\frac{\pi m}{16 \hbar^3} \int dk_1 k_1^3 z e^{-\beta \lambda} 
e^{-\beta \hbar^2 k_1^2/2m} 
\int d\Omega_1 d\Omega_2 e^{-\frac{\hbar k_1^2}{m \omega_{\rm dip}} 
(1-\cos \alpha)}
\end{equation}
where $\alpha$ is the angle between the two vectors ${\vec k}_1$ and
${\vec k}_2$ corresponding to the two scattered non-condensed particles.
Since the condensate in the dip is very localized, its typical linear
dimension being $l_{\rm dip} = \sqrt{\hbar/m \omega_{\rm dip}}$, then
$l_{\rm dip} \ll l_{\rm thermal}$, where $l_{\rm thermal}=\sqrt{\hbar^2/2
m k_{\rm B} T}$ is the thermal de Broglie wavelength. This means that in
the above expression the first (thermal)  exponential factor dominates,
and we can approximate $d_3$ by

\begin{equation}
d_3 \approx - \frac{\pi m (4 \pi)^2}{16 \hbar^3} \int dk_1 k_1^3 
z e^{-\beta \lambda} e^{-\beta \hbar^2 k_1^2/ 2m} = 
- \frac{\pi^2 m^2}{4 \hbar^4} \rho_{\rm E} v_T
\end{equation}
where $v_T = \sqrt{ 2 k_{\rm B} T/m}$ is the thermal velocity of
non-condensed particles, and $\rho_{\rm E}= N_{\rm E}/V = V^{-1}
\sum_{\vec k} z e^{-\beta \lambda} e^{-\beta \hbar^2 k_1^2/ 2m}$ is the
density of non-condensed particles in the antisymmetric states.

Given the expression for $d_3$, we can give an estimate for the
decoherence rate. In the $|J,M \rangle$ basis, with $J=N/2$ and $J_z |J,M
\rangle = (1/2) (N_A - N_B) |J, M \rangle$, the master equation reads

\begin{equation}
\dot{\rho}(M,M') \approx 16 v^2 d_3 (M-M')^2 \rho(M,M') =
16 \pi^3 \gamma (M-M')^2 \rho(M,M') 
\end{equation}
For the quantum superposition state $|N,0\rangle + |0,N\rangle$, $M-M'=N$,
so our final estimate for the decoherence rate is

\begin{equation}
t_{\rm dec}^{-1} \approx 16 \pi^3 \left(
4 \pi a^2 \frac{N_{\rm E}}{V} v_T \right) N^2
\end{equation}

  The only $O(z)$ contribution to the amplitude decoherence can come from
a term proportional to $d_4$. However, under closer inspection $d_4$ turns
out to be $O(z^2)$. $d_4$ comes from a depletion/growth inelastic process
such that a noncondensate particle in the initial state of $s_1$ collides
with a condensate particle and as a result they both end in noncondensate
modes $s_2$ and $s_3$. For this to happen $s_1$ must have a sufficiently
high energy to overcome a gap between the condensate mode and the
environmental modes. Thus in this case $n_{s_1}^{O}=O(z^2)$ and not
$O(z)$.

\section{Discussion}

The aim of this paper was to discuss the ``longevity'' of Schr\"odinger
cats in BEC's. We have shown that while in the standard traps decoherence
rates are significant enough to prevent long-lived macroscopic superpositions
of internal states of the condensate, the strategy of trap engineering and
symmetrization of the
environment will be able to deal with that issue. 

What remains to be considered is how  one can generate such macroscopic quantum
superposition, and how one can detect it. The issue of generation was
already touched upon in Refs. \cite{lz,gs}. We have little to add to this.
However, in the context of the Gordon and Savage proposal, it is fairly
clear that the time needed to generate the cat state would have to be short
compared to the decoherence time. If our estimates of Eq.(\ref{dr}) are 
correct, symmetrization procedure appears necessary for the success of such 
schemes.

Detection of Schr\"odinger cat states is perhaps a more challenging subject.
In principle, states of the form $( |N,0\rangle + |0,N\rangle) /\sqrt{2}$
have a character of GHZ states, and one could envision performing 
measurements analogous to those suggested in \cite{molmer} and
carried out in \cite{four}, where a 4-atom entangled state was studied.
However, this sort of parity-check strategy, appropriate for $N\leq 10$,
is likely to fail when $N$ is larger, or when (as would be the case for
the ``quasi-squeezed'' states anticipated here \cite{moon}) $N$ is not
even well defined. 

A strong circumstantial evidence can be nevertheless obtained from two
measurements. The first one would consist of a preparation of the cat state,
and of a measurement of the internal states of the atoms. It is expected that
in each instance all (to within the experimental error) would turn up to be
in either A or B states. However, averaged over many runs, the number of either
of these two alternatives would be approximately equal. Decoherence in which
the environment also ``monitors'' the internal state of the atoms in the
A versus B basis would not influence this prediction. We need to check
separately whether the cat state was indeed coherent. To do this, one could
evolve the system ``backwards''. However, this is not really necessary. For, 
as Gordon and Savage point out, 
when, in their scheme,
we let the system evolve unitarily
for more or
less twice
the time needed for the generation of the cat state, it will approximately
return to the initial configuration. Thus, we can acquire strong evidence
of the coherence of the cat provided that this 
unitary
return to the initial
configuration can be experimentally confirmed.

These are admittedly rather vague ideas, which serve more as a ``proof of
principle'' rather than as a blueprint for an experiment. Nevertheless,
they may, we hope, encourage more concrete investigation of 
such issues with a specific experiment in mind.

\section{Acknowledgements} 

We are indebted to E.Cornell, R.Onofrio, E.Timmermans, and especially to
J.Anglin for very useful comments. This research was supported in part by
NSA.

\end{document}